\documentclass[a4paper,fleqn,usenatbib]{mnras}

\usepackage{newtxtext,newtxmath}

\usepackage[T1]{fontenc}
\usepackage{ae,aecompl}

\usepackage{graphicx}	
\usepackage{amsmath}	
\usepackage{amssymb}	

\title[The battle of entropy]{Mirror principle and the red-giant bump: the battle of entropy in low-mass stars}

\author[S. Hekker et al.]{
S. Hekker$^{1,2}$\thanks{E-mail: hekker@mps.mpg.de},
G.C. Angelou$^{3}$,
Y. Elsworth$^{4,2}$
and S. Basu$^{5}$
\\
$^{1}$Max Planck Institute for Solar System Research, Justus-von-Liebig Weg 3, D-37077 G\"ottingen, Germany\\
$^{2}$Stellar Astrophysics Centre, Department of Physics and Astronomy, Ny Munkegade 120, DK-8000 Aarhus C, Denmark\\
$^{3}$Max Planck Institute for Astrophysics, Karl-Schwarzschild-Str. 1, D-85741 Garching, Germany\\
$^{4}$School of Physics and Astronomy, University of Birmingham, Birmingham B15 2TT, UK\\
$^{5}$Department of Astronomy, Yale University, New Haven, CT 06520, USA
}

\date{Accepted XXX. Received YYY; in original form ZZZ}

\pubyear{2018}

\begin{document}
\label{firstpage}
\pagerange{\pageref{firstpage}--\pageref{lastpage}}
\maketitle

\begin{abstract}
The evolution of low-mass stars into red giants is still poorly understood. During this evolution the core of the star contracts and, simultaneously, the envelope expands -- a process known as the `mirror'. Additionally, there is a short phase where the trend for increasing luminosity is reversed. This is known as the red-giant-branch bump. We explore the underlying physical reasons for these two phenomena by considering the specific entropy distribution in the star and its temporal changes. We find that between the luminosity maximum and luminosity minimum of the bump there is no mirror present and the star is fully contracting. The contraction is halted and the star regains its mirror when the hydrogen-burning shell reaches the mean molecular weight discontinuity. This marks the luminosity minimum of the bump.
\end{abstract}




\section{Introduction}
\begin{figure}
\centering
\includegraphics[width=\linewidth]{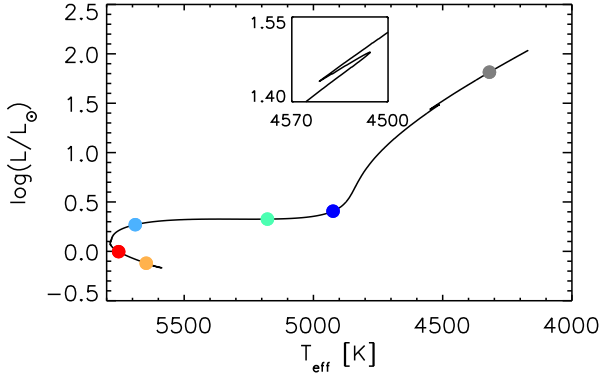}
\caption{Herzsprung-Russell diagram (H-R diagram) of a 1\,M$_{\odot}$ track with solar composition computed with MESA. The coloured dots indicate models for which properties are shown in Figs~\ref{entropy} and \ref{epsgprof}. The inset shows a zoom of the red-giant-branch bump (RGBB).}
\label{HRD}
\end{figure}

The post-main sequence evolution of low-mass stars into red-giant stars is of fundamental interest for understanding stellar evolution, as well as for the study of galactic evolution. Yet at the same time it remains an enigma. We are particularly interested in two phenomena seen in this evolutionary progression. 
Firstly, there is the expansion of the envelope and simultaneous contraction of the inner regions of low-mass stars in the subgiant phase and up to the tip of the red-giant branch (RGB), known as the mirror phenomenon. During this time mass shells in the envelope move outwards and mass shells in the inner regions move inwards.
Secondly, there is the red-giant-branch bump (RGBB) which is the phase where the trend in increasing luminosity reverses for a short time before it increases again, causing a zig-zag in the evolutionary track (inset of Fig.~\ref{HRD}). These are well known features in stellar models. The RGBB is clearly seen in observations of, for instance, open and globular clusters. The phenomenon causes a star to live longer in a narrow band of luminosities and hence, in an iso-age population, an increased stellar density is observed at the luminosity of the bump. Consequently, the RGBB serves as an important reference point to calibrate models \citep[e.g.][and references therein]{riello2003,angelou2015,joyce2015,khan2018}. Nevertheless, the physics that drives the stellar structure changes associated with the mirror phenomenon and RGBB are not fully understood. 

The reason why stars become red giants, and why red giants exhibit a bump are related. 
Given the importance of understanding red giants, it is not surprising that there have been many attempts to provide a theoretical framework and to give reasons as to why these phenomena occur.
Among these are: the role of the central gravitational field \citep{hoeppner1973,weiss1983}, the effective equation of state \citep{eggleton1991,eggleton1998}, gravothermal instability in the core  \citep{iben1993}, thermal instabilities in the stellar envelope  \citep{renzini1984,renzini1992} and mean molecular weight gradient \citep{stancliffe2009}. None of these previous studies have so far produced clear answers as to why stars become red giants.

Notwithstanding these problems, it is clear that a strong gravitational field and a mean molecular weight gradient play important roles  \citep{stancliffe2009}. For an extensive overview regarding the literature addressing `Why do stars become giants?' we refer the reader to \citet{sugimoto2000} and references therein. These authors  state that `An increase of the entropy in the envelope is indispensable for the evolution to a red giant.' Our current work is motivated by these findings which both directly and indirectly point to the importance of entropy in red-giant branch evolution.

For the RGBB, the naive explanation is that the bump appears when the hydrogen shell burns through the mean molecular weight discontinuity left behind by the deepest extent of the convection zone. At this discontinuity, the amount of hydrogen available for burning increases and consequently there is a re-adjustment of the internal structure. This re-adjustment phase could explain the bump. However, \citet{JCD2015} showed that the burning shell only reaches the mean molecular weight discontinuity at the minimum luminosity. Hence this picture cannot explain the luminosity maximum of the bump completely. Furthermore, it is known that the exact shape of the bump depends on the constituents of the models such as the hydrogen profile, as shown by e.g. \citet{cassisi2002}. 

In this study we link the mirror and the bump to gain more insight in why stars become red giants. We focus particularly on the specific entropy, $s$, \citep[see also][]{avellar2015}  and its temporal gradient. We follow these properties along a 1\,M$_{\odot}$ stellar evolutionary track constructed with solar chemical composition \citep{gs98} computed using MESA  \citep[version r10398,][and references therein]{paxton2018}. We provide explanations for the entropy profiles and the changes in entropy based on physical principles and propose a scenario that sheds light on the mirror and the bump based on these properties.

\section{Specific entropy}
\citet{hansen1994} show that for a constant-composition ideal monoatomic gas (assumed not to be partially ionised) specific entropy can be expressed as:
\begin{equation}
s=\frac{N_{\rm A}k_{\rm B}}{\mu}\ln[T^{5/2}/P]+c,
\label{sstar}
\end{equation}
where $N_{\rm A}$ is Avogadro's constant, $k_{\rm B}$ is Boltzmann's constant, $\mu$ is mean molecular weight, $T$ is temperature, $P$ is pressure and $c$ an integration constant. We omit this constant from consideration in the remainder of this study as we are interested in entropy changes and not in the absolute values.
\newline
The change in specific entropy $ds$ is defined as:
\begin{equation}
ds = \frac{dq}{T},
\label{ds}
\end{equation}
where $T$ is temperature and $dq$ the heat added per unit mass $dq = du + Pdv$ where $u$ is the internal energy, $P$ is pressure and $v = 1/\rho$ is specific volume, with $\rho$ the density.
By expressing $u$ and $\rho$ in terms of $P$ and $T$ and expanding them out into partials with respect to $P$ and $T$, and subsequently transform them using standard thermodynamic rules \citet{hansen1994} arrive at:
\begin{equation}
\frac{ds}{dr} = c_{\rm P}(\nabla-\nabla_{\rm ad})\frac{d\ln P}{dr},
\label{dsdr}
\end{equation}
with 
\begin{equation}
\nabla=\frac{d \ln T}{d \ln P},
\end{equation}
$\nabla_{\rm ad}$ the adiabatic temperature gradient with pressure, $r$ the radius ordinate, and $c_{\rm P}$ the specific heat at constant pressure.

\begin{figure*}
\centering
\includegraphics[width=\linewidth]{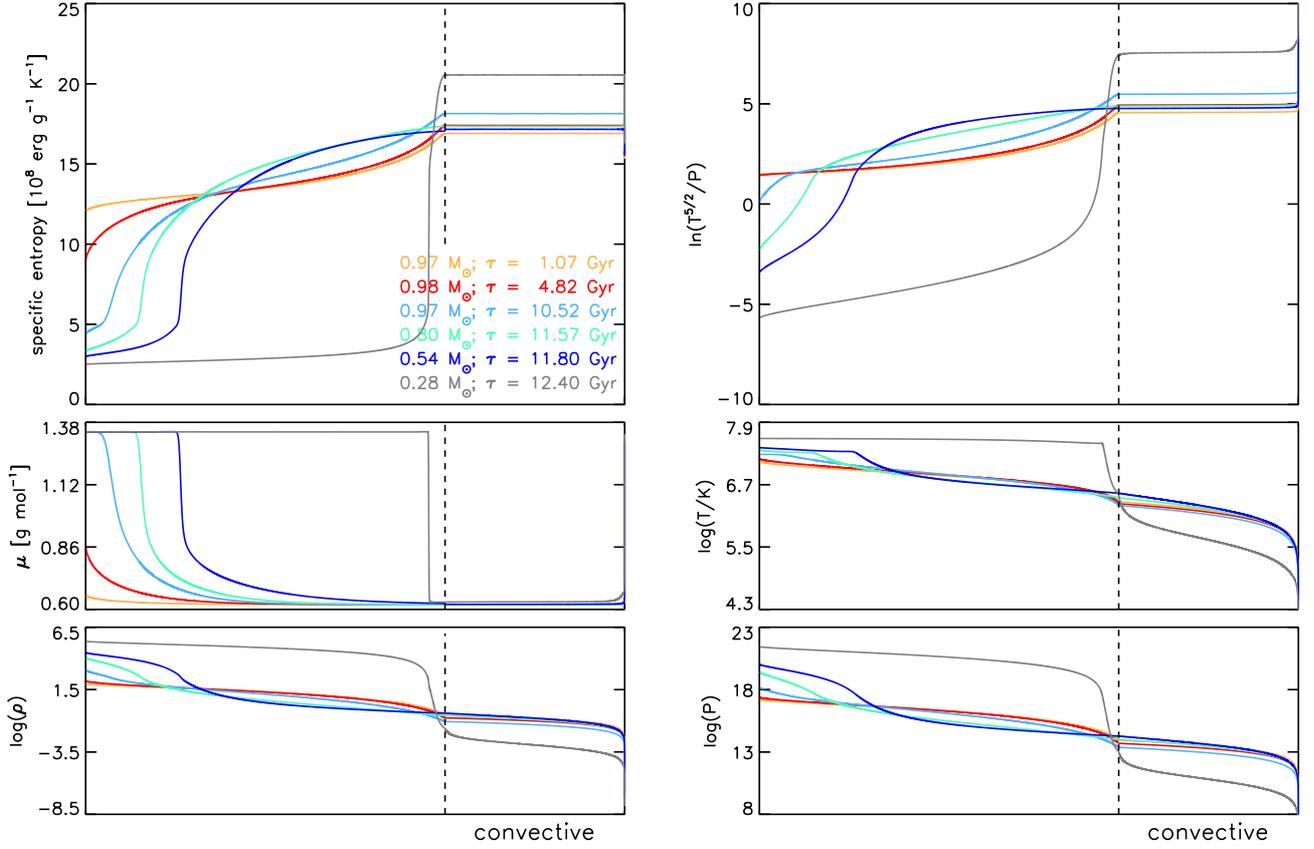}
\caption{\textsc{top left}: specific entropy profiles as a function of mass ordinate of six 1\,M$_{\odot}$ models at different ages. For visualisation purposes the model structures are divided in a non-convective region and a convective region, where within a region the mass varies linearly. The base of the convection zone is indicated with a vertical dashed line. The (fractional) masses of each model at the dashed line is indicated together with the age ($\tau$) of the model. The total range of the mass ordinate is 0 to 1\,M$_{\odot}$. The evolutionary phases of the curves are indicated in Fig.~\ref{HRD} with dots of corresponding colours. \textsc{top right}: $\ln(T^{5/2}/P)$ as a function of mass ordinate; \textsc{middle left}: mean molecular weight as a function of mass ordinate; \textsc{middle right}: logarithmic temperature as a function of mass ordinate; \textsc{bottom left}: logarithmic density as a function of mass ordinate; \textsc{bottom right}: logarithmic pressure as a function of mass ordinate. The colours and dashed lines have the same meaning as in the top left panel.}
\label{entropy}
\end{figure*}

Here, we investigate models along a 1\,M$_{\odot}$ stellar evolutionary track with solar composition (see Fig.~\ref{HRD}) computed with MESA\footnote{We use MESA with the maximum and minimum size of a (fractional) mass shell to be between $10^{-4}$ and $10^{-18}$, i.e. we set the MESA options max\_dq = $10^{-4}$ and min\_dq=$10^{-18}$. This results in about 14\,600 meshpoints across the star, and the number of meshpoints varies slowly. This is important to obtain most accurate time derivatives. For the same reason, we use small timesteps of $10^6$ yr on the red-giant branch and $10^7$ yr on the main sequence. Finally, we use strict convergence criteria for optimal results.}. In Fig.~\ref{entropy}, we show the entropy profiles as obtained from MESA, i.e. using the EOS tables (see Section 3), together with mean molecular weight, density, temperature, pressure and $T^{5/2}/P$ profiles for models indicated with the coloured dots in Fig.~\ref{HRD}. Note that we chose to show the profiles for models that are either before the first dredge-up or after the bump so as to not be influenced by the mean molecular weight discontinuity which will be addressed in section~\ref{sect:bump}.
We find that the specific entropy decreases with time at the location where fusion is dominating, i.e. either in the core or a shell around the core depending on the evolutionary phase of the star. This decrease in specific entropy is related to the increase in the mean molecular weight (middle left panel of Fig.~\ref{entropy}) and decrease of $T^{5/2}/P$ (see, Eq.~\ref{sstar} and top right panel of Fig.~\ref{entropy}) in the core. 
Surrounding the burning core is a radiative layer. In this layer, hydrostatic equilibrium requires that $d \ln P / dr \leq 0$ and $\nabla < \nabla_{\rm ad}$, thus $ds/dr >0$ and the specific entropy increases outwards (see Eq.~\ref{dsdr}). 
For low-mass stars, a convective layer is present on top of the radiative region. In this convective region the composition is well mixed and uniform, and convection is to good approximation isentropic (except in super-adiabatic layers, which we ignore here), i.e. $\nabla = \nabla_{\rm ad}$. As per Eq.~\ref{dsdr}, this leads to a uniform specific entropy and mean molecular weight profile across the convection zone. 
 
\begin{figure}
\centering
\begin{minipage}{0.95\linewidth}
\includegraphics[width=\linewidth]{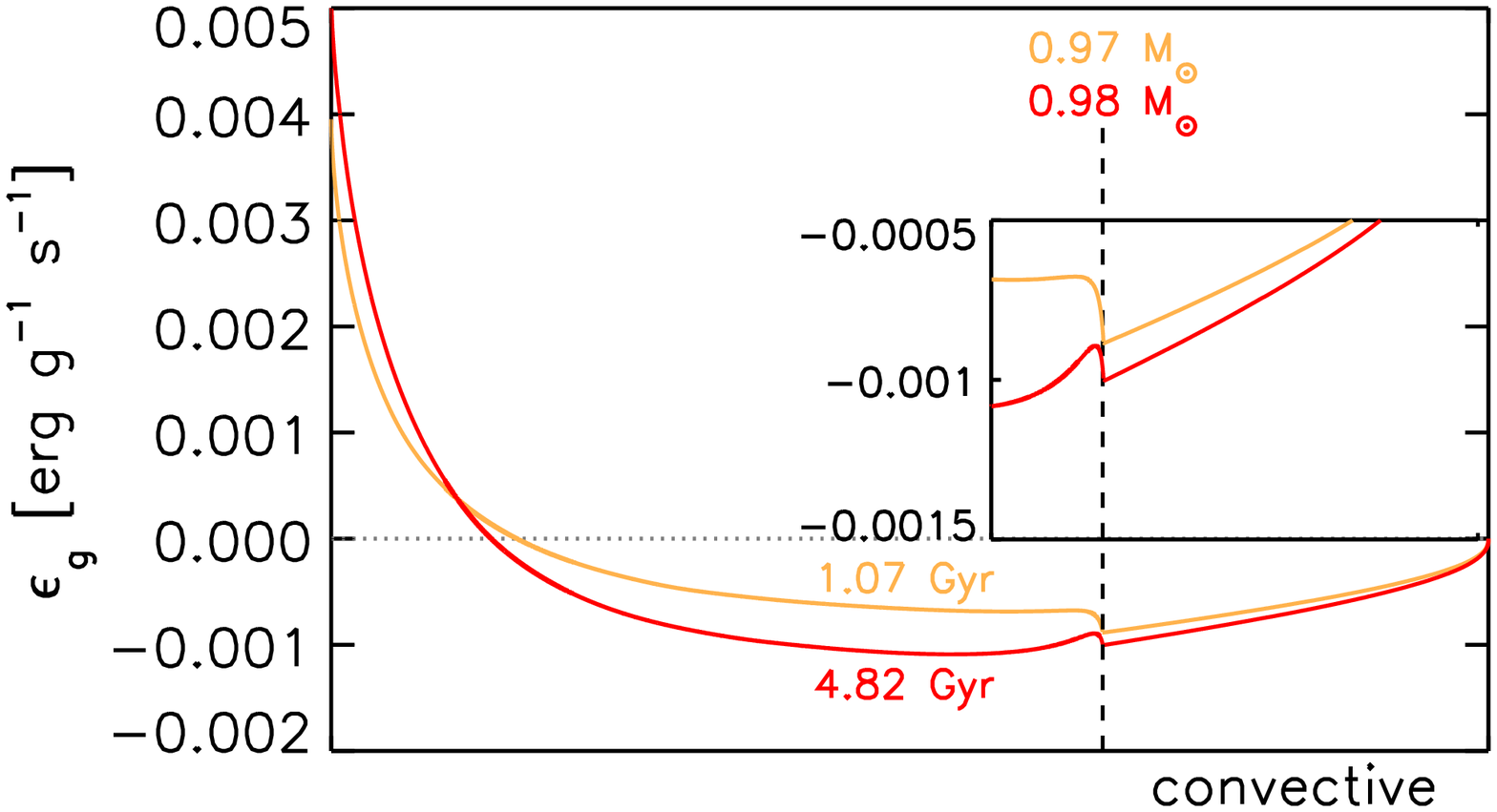}
\end{minipage}
\begin{minipage}{0.95\linewidth}
\includegraphics[width=\linewidth]{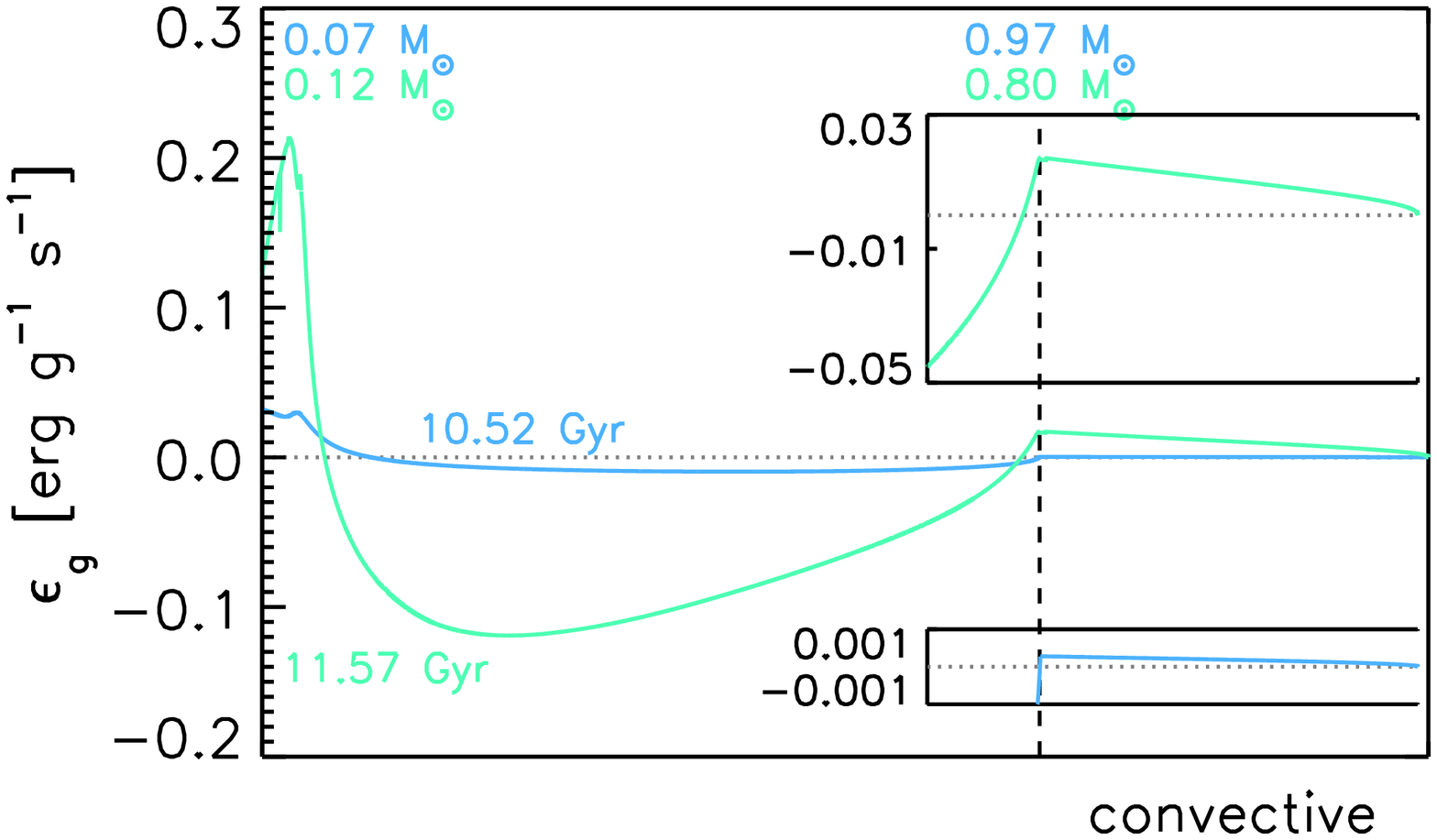}
\end{minipage}
\begin{minipage}{0.95\linewidth}
\includegraphics[width=\linewidth]{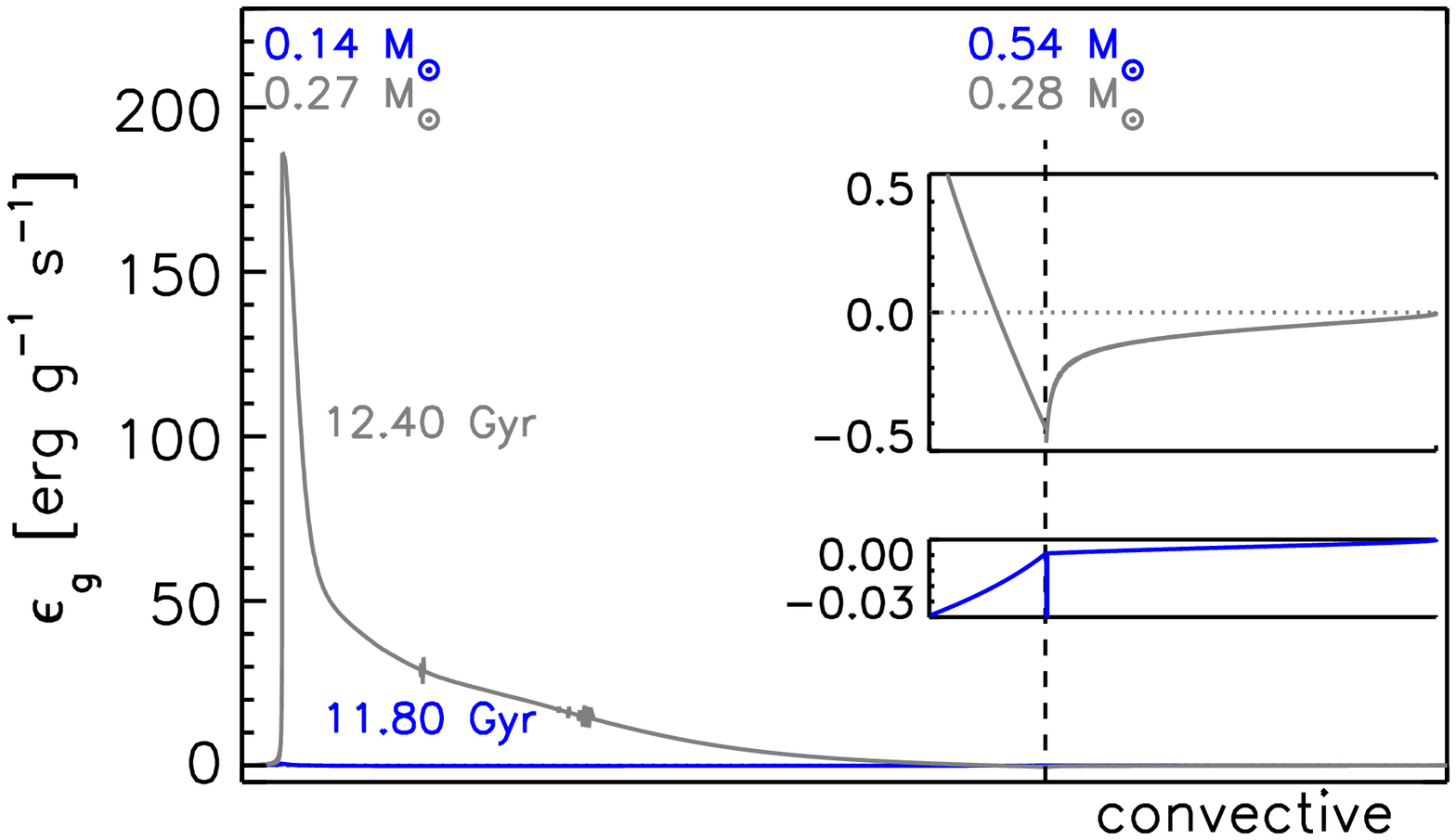}
\end{minipage}
\caption{Profiles of $\epsilon_{\rm g}$ as a function of mass ordinate for stars on the main sequence (\textsc{top}), stars in the subgiant phase (\textsc{middle}) and stars on the red-giant branch (\textsc{bottom}). For comparison reasons the base of the convection zone is scaled to be at the vertical dashed lines in all panels. The mass of each model at the base of the convection zone is indicated. For stars with shell burning (middle and bottom panels) the peaks in $\epsilon_{\rm g}$ are also aligned with the mass at the location of maximum $\epsilon_{\rm g}$ of each model indicated in the top left corner of the panel. The mass varies linearly within the different regions with the total range of the mass ordinate ranging from 0 to 1\,M$_{\odot}$. The insets show zooms of the profiles ranging from just below the base of the convection zone to the surface. Note that the irregular behaviour in some parts of the $\epsilon_{\rm g}$ profiles in the bottom panel is due to numerical errors in the time derivatives.}
\label{epsgprof}
\end{figure}
 
\section{Rate of change of specific entropy}
In stars, the rate of change of specific entropy is proportional to $\epsilon_{\rm g}$, the `gravothermal' energy generation rate, which is defined as \citep{kippenhahn2012,iben2013}:
\begin{equation}
    \epsilon_{\rm g} = -T\frac{\partial s}{\partial t} = - \frac{\partial u}{\partial t} +\frac{P}{\rho^2}\frac{\partial \rho}{\partial t} = \epsilon_{\rm internal}+ \epsilon_{\rm compression},  
  \label{eq_epsg} 
\end{equation}
where $\epsilon_{\rm internal} = - \partial u / \partial t$ is the rate of change of the internal energy per unit mass, $\epsilon_{\rm compression} = (P / \rho^2) (\partial \rho / \partial t)$ is the rate at which work is being done per unit mass to compress matter, and $t$ is time. So in regions in the star where $\epsilon_{\rm compression} >0$ the density increases with time. We note here that in this formulation the changes in particle number abundances due to nuclear transformations and mixing mechanisms are inherently taken into account.\\
\newline
In MESA the specific entropy is calculated from the 2005 update of the OPAL EOS tables \citep{rogers2002}, and the table sets the zero point of the specific entropy. For the current work the exact value of the specific entropy does not matter, what matters are the specific entropy differences. The value of $\epsilon_{\rm g}$ is computed in MESA as \citep[eq. 12 in][]{paxton2011}:
\begin{equation}
\epsilon_{\rm g} = -Tc_P\left[ (1-\chi_T\nabla_{\rm ad})\frac{d\ln T}{dt}-\chi_\rho\nabla_{\rm ad}\frac{d \ln \rho}{dt}\right],
\label{epsgMESA}
\end{equation}
where $d \ln T/ dt$ and $d \ln \rho/ dt$ are Langrangian time derivatives, and
\begin{eqnarray}
\chi_\rho=\frac{\partial\ln P}{\partial \ln \rho}\bigg|_T~~~~~~~~
\chi_T=\frac{\partial\ln P}{\partial \ln T}\bigg|_\rho.
\end{eqnarray}
This formulation in MESA is equivalent to the formulation presented in Eq.~4.47 of \citet{kippenhahn2012}:
\begin{equation}
\epsilon_{\rm g} = -c_P\frac{\partial T}{\partial t}+\frac{\delta}{\rho}\frac{\partial P}{\partial t} = -c_P T\left(\frac{1}{T}\frac{\partial T}{\partial t} -\frac{\nabla_{\rm ad}}{P}\frac{\partial P}{\partial t} \right),
\end{equation}
with
\begin{equation}
\delta = -\left(\frac{\partial \ln \rho}{\partial \ln T} \right)_P,
\end{equation}
which is in turn equivalent to Eq.~\ref{eq_epsg}.

We now show $\epsilon_{\rm g}$ profiles as described by Eq.~\ref{epsgMESA} as a function of mass ordinate of 1\,M$_{\odot}$ models at different ages in Fig.~\ref{epsgprof}. We find that in each of these $\epsilon_{\rm g}$ profiles there is at least one sign change. Both models in the middle plot actually show a change in sign of $\epsilon_{\rm g}$ twice, one in the deep interior just outside the maximum value of $\epsilon_{\rm g}$, and one just below the base of the convection zone. In the next section we discuss the changes in sign of $\epsilon_{\rm g}$ in the context of the attributes of the mirror phenomenon.

\begin{figure*}
\centering
\begin{minipage}{0.48\linewidth}
\includegraphics[width=\linewidth]{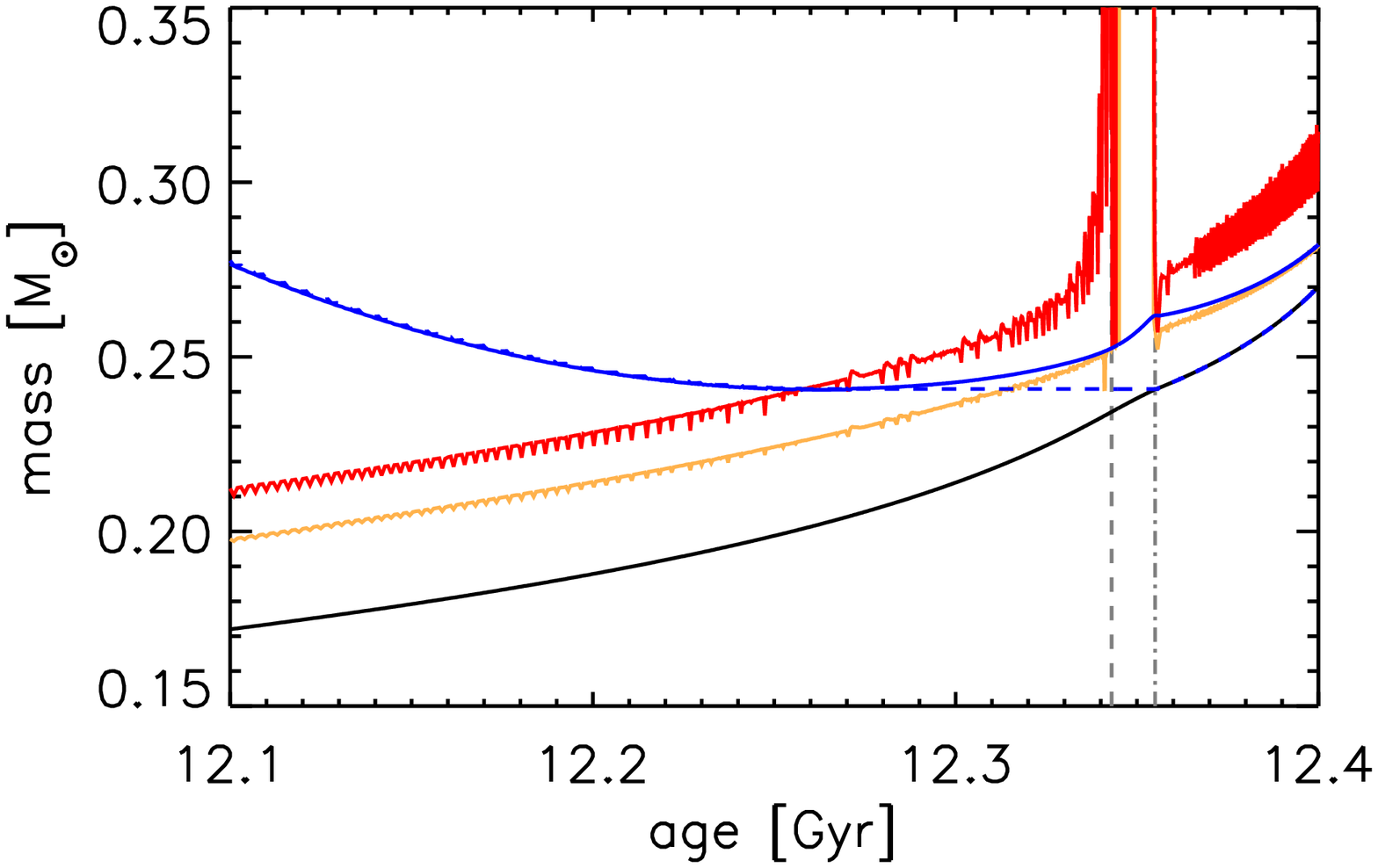}
\end{minipage}
\begin{minipage}{0.48\linewidth}
\includegraphics[width=\linewidth]{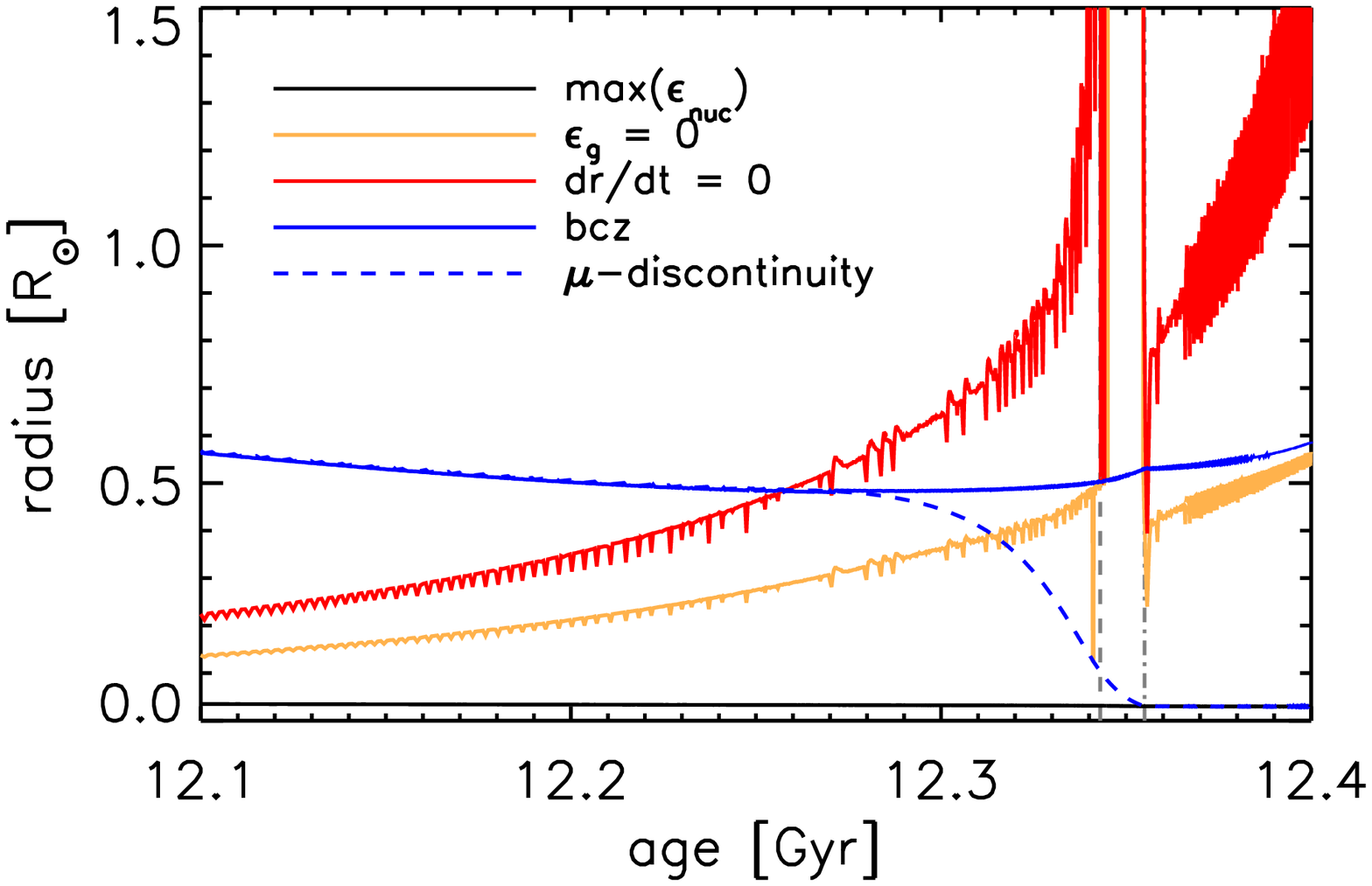}
\end{minipage}
\begin{minipage}{0.48\linewidth}
\includegraphics[width=\linewidth]{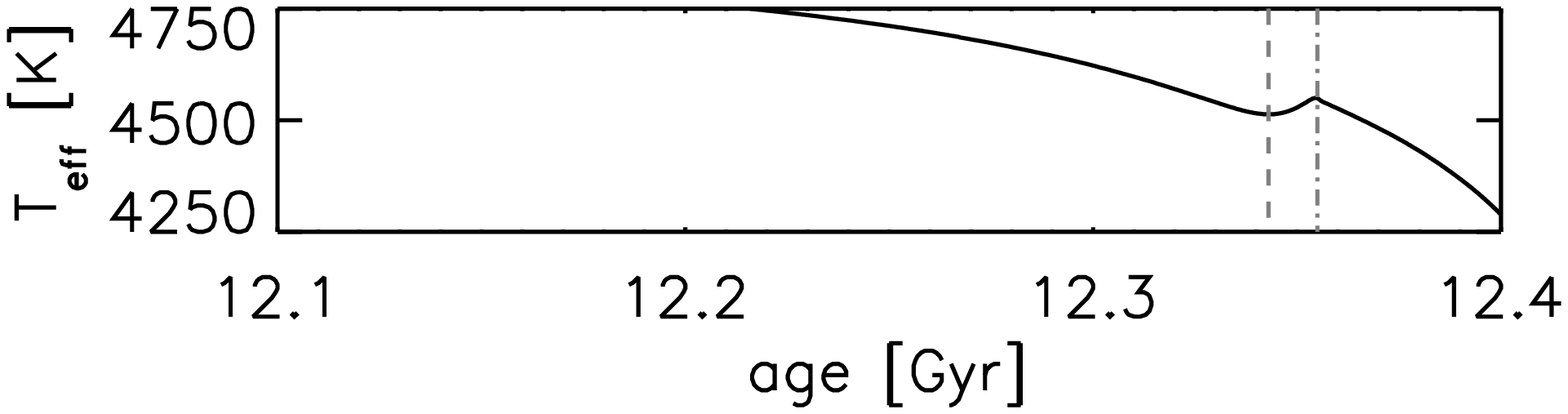}
\end{minipage}
\begin{minipage}{0.48\linewidth}
\includegraphics[width=\linewidth]{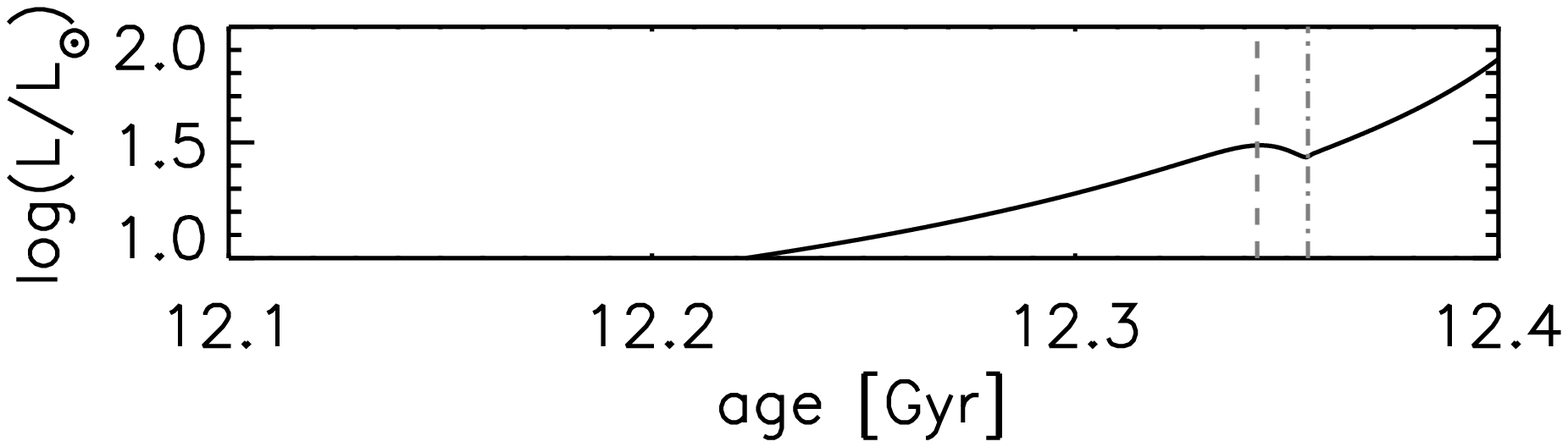}
\end{minipage}
\caption{\textsc{top}: the evolution around the bump of the location of the base of the convection zone (blue solid line), the stationary point ($\partial r/\partial t = 0$, red), the pivot ($\epsilon_{\rm g} = 0$, orange), the peak of the burning (max($\epsilon_{\rm nuc}$), black) and the mean molecular weight discontinuity (blue dashed line) as function of mass ordinate (\textsc{left}) and radius ordinate (\textsc{right}). \textsc{bottom}: effective temperature (\textsc{left}) and luminosity (\textsc{right}) as a function of age. The vertical grey dashed and dashed-dotted lines indicate the age of the maximum luminosity and minimum luminosity of the bump feature, respectively. As in Fig.~\ref{epsgprof} the irregular behaviour in the  $\epsilon_{\rm g}$=0 and $\partial r/\partial t=0$ profiles in the central panels is due to numerical effects in the time derivatives.}
\label{location}
\end{figure*}

\section{Attributes of the mirror phenomenon}
The mirror phenomenon is a well-known phenomenon seen in stars with shell-burning sources such as subgiants and red-giant stars. In these evolutionary phases the mirror phenomenon involves the expansion of the envelope of the star, while at the same time the inner regions contract. Here we introduce two well-defined attributes of the mirror: a `stationary point' and a `pivot'.

A `stationary point' indicates a location below which shells move inwards and above which shells move outwards (or vice versa). It is defined as the mass shell where the derivative of the radius ($r$) with respect to time ($t$) is zero:
\begin{equation}
\partial r/\partial t = 0.
\label{stationary}
\end{equation}

A `pivot' is defined here as the location where the gravothermal energy generation rate changes sign, i.e.
\begin{equation}
\epsilon_{\rm g} =  0.
\label{mirror}
\end{equation}
This definition of the pivot can be interpreted as the location where shells transition from compressing (increasing density, in which case the released gravitational energy results in an increase in the internal energy) to expanding \citep[decreasing density, where internal energy is used up in order to expand,][]{iben2013}. Note that the pivot as defined here also exist on the main sequence (see top panel of Fig.~\ref{epsgprof}).

As the pivot is located where the temporal variation of entropy changes sign, we deduce from Eq.~\ref{sstar} that either $\mu$ or $T^{5/2}/P$ decreases on one side of the pivot and simultaneously increases on the other side of the pivot. Temporal variation of $T^{5/2}/P$ seem to be the dominant cause of the pivot, as $T^{5/2}/P$ decreases with time in the inner regions and increases with time in the outer envelope of an evolving star, as shown in the top right panel of Fig.~\ref{entropy}. The  temporal variations of  the mean molecular weight do not show such changes (see middle left panel of Fig.~\ref{entropy}). 

We emphasise here that the convective region is isentropic and consequently $\partial{s}/ \partial{t}$ is the same in all parts of the convective region. As the temperature is always positive, a uniform $\partial{s}/\partial{t}$ in spatial terms implies that $\epsilon_{\rm g}$ cannot change sign in the convective region (see Eq.~\ref{eq_epsg}), and hence there cannot be a pivot in a convective region.

\section{The bump}
\label{sect:bump}
In this section we investigate the bump, i.e. the short phase of contraction and brief reversal in the increasing luminosity along the RGB. More specifically, we investigate the impact of the mean molecular weight discontinuity on the entropy and how this could explain the bump. To do so, we first investigate the evolution of the pivot and stationary point, in the phase before the luminosity maximum, between the luminosity maximum and the luminosity minimum and shortly after the luminosity minimum. We subsequently perform a comparison between an evolutionary track with a mean molecular weight discontinuity and a track where the mean molecular weight discontinuity has been artificially removed.

We  look at the red-giant-branch bump of a 1\,M$_{\odot}$ evolutionary track with solar metallicity. We follow the location of the pivot, i.e. $\epsilon_{\rm g} = 0$ (Eq.~\ref{mirror}), as well as the location of the stationary point, i.e. $\partial r/\partial t = 0$ (Eq.~\ref{stationary}), as a function of time (see Fig.~\ref{location}). These locations change as stars evolve. The ages presented here are specific to the chosen evolutionary track. These same phenomena will appear at different ages for tracks of different masses and physics. In this stage of the evolution, the density in the core increases and the hydrogen-burning shell moves to larger fractional mass while it remains approximately at constant fractional radius. As the burning shell is the primary source of energy, it dictates the global behaviour of the pivot and the stationary point. As the burning shell advances in mass the pivot and stationary point also move to higher fractional mass. The stationary point crosses the inward moving base of the convection zone at an age of about 12.26 Gyr in this particular stellar evolution track (red crossing blue in the top panels of Fig.~\ref{location}). 

\begin{figure}
\centering
\begin{minipage}{\linewidth}
\includegraphics[width=\linewidth]{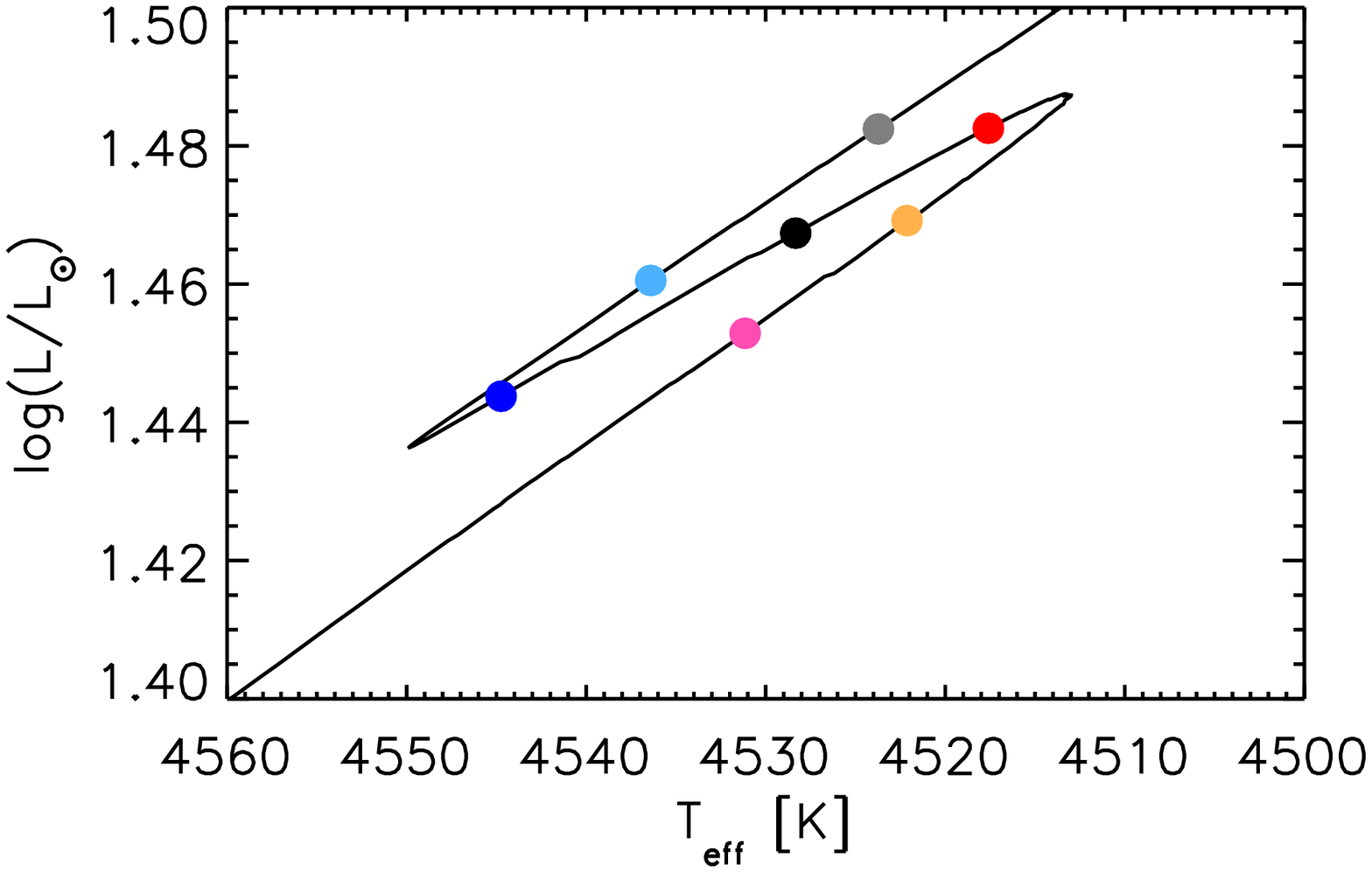}
\end{minipage}
\begin{minipage}{\linewidth}
\includegraphics[width=\linewidth]{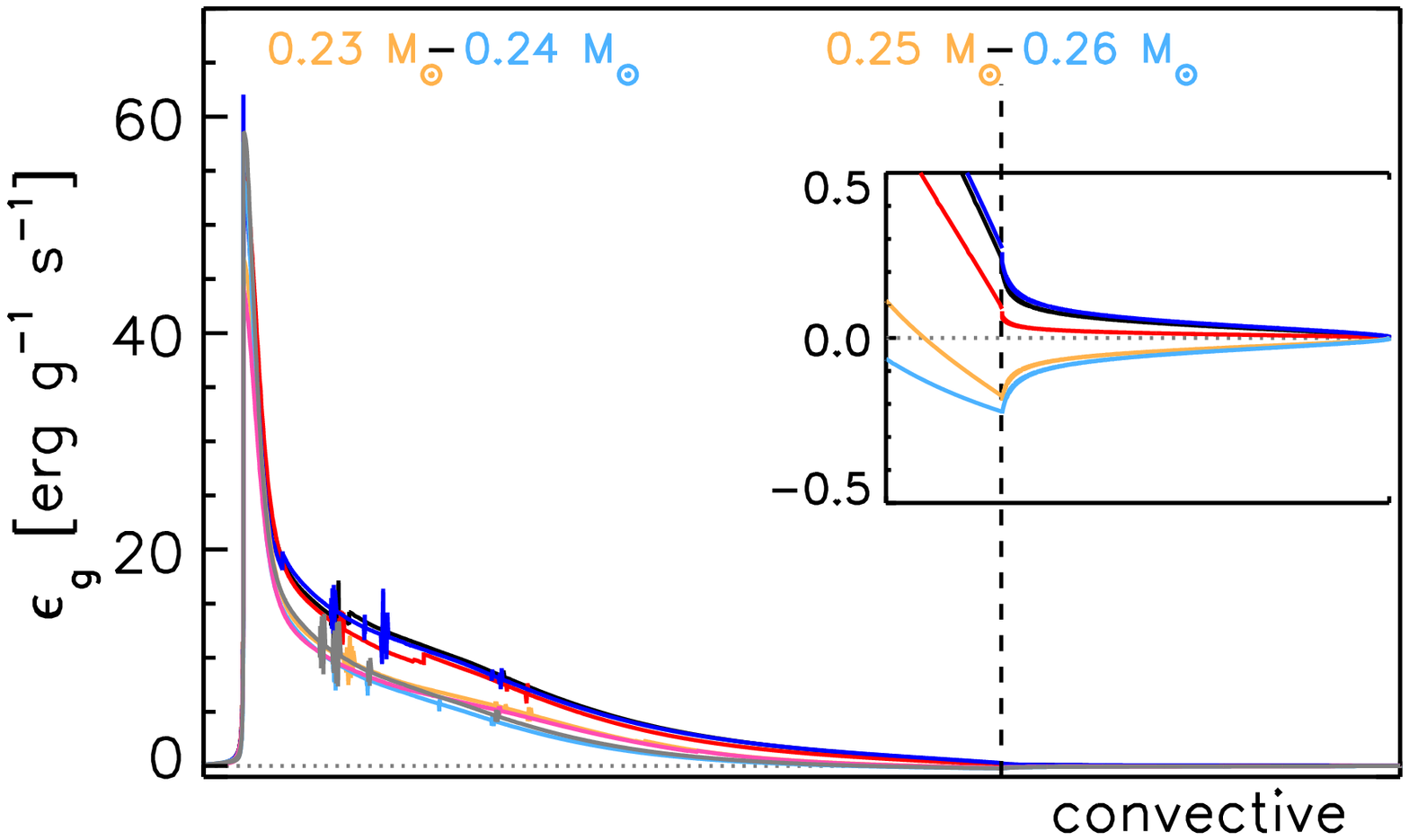}
\end{minipage}
\caption{The bump feature in the H-R diagram (\textsc{top}) with the dots indicating models for which the $\epsilon_{\rm g}$ profiles (\textsc{bottom}) are shown in a similar way as in Fig.~\ref{epsgprof}. In the bottom panel the inset shows details of the behaviour of $\epsilon_{\rm g}$ in the convection zone (note that blue and black nearly overlap in the convection zone). The order of the models in time is pink, orange, red, black, blue, light blue, grey (the pink and grey model are not shown in the inset as these would overlap with the orange and light blue models). See text for more details. Note that the irregular behaviour in some parts of the $\epsilon_{\rm g}$ profiles in the bottom panel is due to numerical errors in the time derivatives.}
\label{bump}
\end{figure}

Subsequently, at an age of about 12.344 Gyr for this particular stellar evolution track, the location where $\epsilon_{\rm g} = 0$ reaches the base of the convection zone.  At this point, the whole radiative region is undergoing compression. We note that this coincides with the luminosity maximum of the bump (see bottom panels of Fig.~\ref{location}). Since the entropy over the convection zone is uniform, neither $\partial s / \partial t$ nor $\epsilon_{\rm g}$ changes sign within the convection zone (see Section 4). To further illuminate this point, we show the $\epsilon_{\rm g}$ profiles of some models in the vicinity of the bump in Fig.~\ref{bump}. These profiles show that there is no longer a pivot between the hydrogen-burning shell and the base of the convection zone for models between the luminosity maximum and luminosity minimum (red, black and dark blue curves in Fig.~\ref{bump}).

\begin{figure}
\centering
\includegraphics[width=\linewidth]{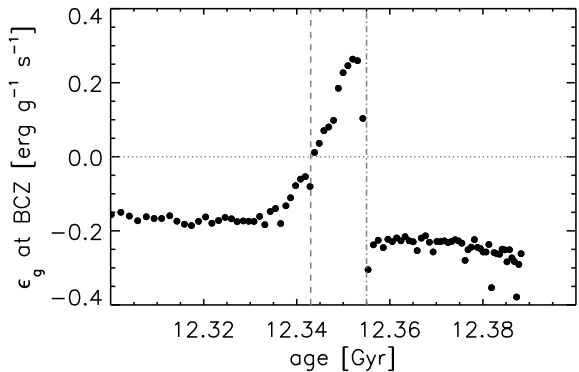}
\caption{The value of $\epsilon_{\rm g}$ at the base of the convection zone as a function of age. The vertical dashed and dashed-dotted lines indicate the stellar ages of the luminosity maximum and luminosity minimum of the bump, respectively. The horizontal dotted line indicates zero.}
\label{bcz}
\end{figure}

In Fig.~\ref{bcz} we present the value of $\epsilon_{\rm g}$ at the base of the convection zone for all computed models around the age where the RGB bump is observed. This figure shows that not only does the luminosity maximum coincide with the start of the phase where $\epsilon_{\rm g}$ is positive at the base of the convection zone, but also the luminosity minimum that follows coincides with the transition of $\epsilon_{\rm g}$ to a negative value. Put another way, across the bump (between the maximum and minimum luminosity) $\epsilon_{\rm g}$ is positive throughout the stellar model and there is no pivot present: the entire stellar model contracts. 
This contraction halts when the hydrogen shell reaches the mean molecular weight discontinuity left behind by the receding convection zone \citep[in line with the results by][]{JCD2015} at an age of about 12.355 Gyr for this particular evolutionary track. At this point the extra hydrogen fuels an increase in the nuclear burning and the model regains its pivot and continues to ascend the RGB. 

\begin{figure}
\centering
\includegraphics[width=\linewidth]{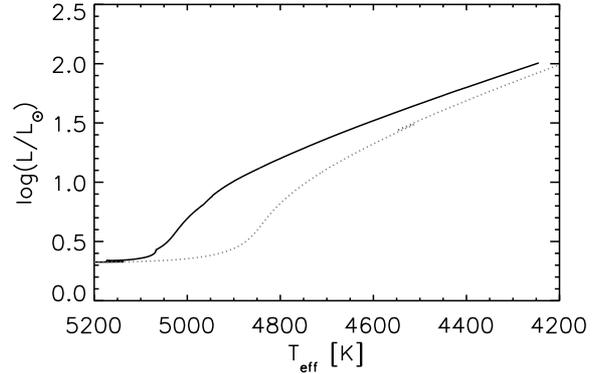}
\caption{Stellar evolutionary track of a standard 1\,M$_{\odot}$ stellar evolutionary track with a mean molecular weight discontinuity (dotted line) and with the mean molecular weight discontinuity removed (solid line, see text for more details).}
\label{HRDnodmu}
\end{figure}

To further understand why the stellar model loses its pivot at the luminosity maximum of the bump, we investigate the temporal changes in $T^{5/2}/P$, the entropy and $\epsilon_{\rm g}$. We first look at the models in which the mean molecular weight discontinuity has been artificially removed. In a given model, we achieved this by homogenising $\mu$ and the relevant mass fractions down to the midpoint between the hydrogen-burning shell and the base of the convection zone. We let this model evolve for one timestep and apply the same process to the resulting model. We continue to apply this process till the track reaches a stage beyond which the bump could be expected. 
The resulting stellar evolution track along the red-giant branch together with the standard 1\,M$_{\odot}$ stellar evolution track are shown in Fig.~\ref{HRDnodmu}. As expected, the track of the modified models follows a different path from the standard one. More importantly, the track without a mean molecular weight discontinuity does not contain a bump.

\begin{figure}
\centering
\begin{minipage}{0.9\linewidth}
\includegraphics[width=\linewidth]{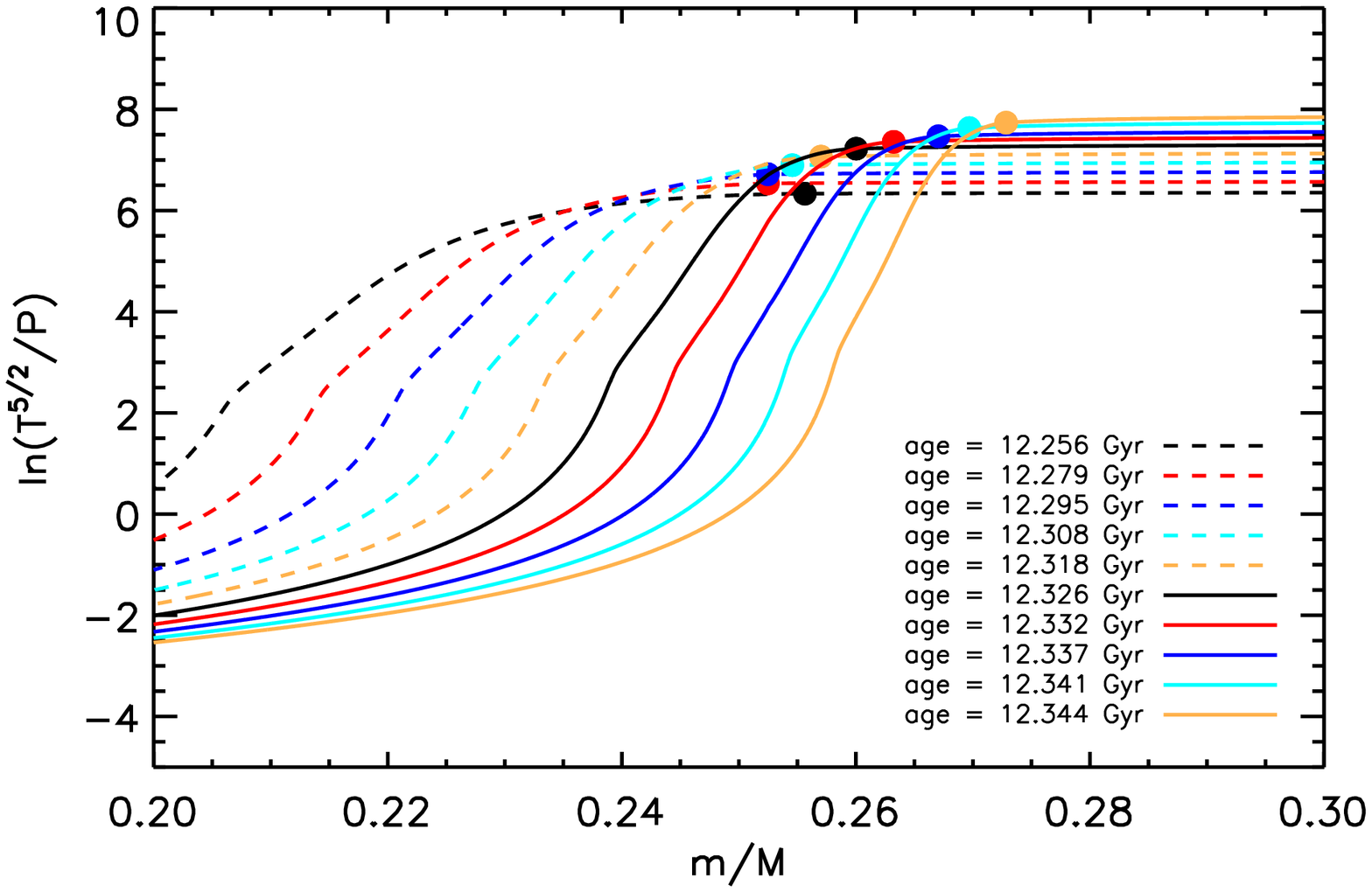}
\end{minipage}
\begin{minipage}{0.9\linewidth}
\includegraphics[width=\linewidth]{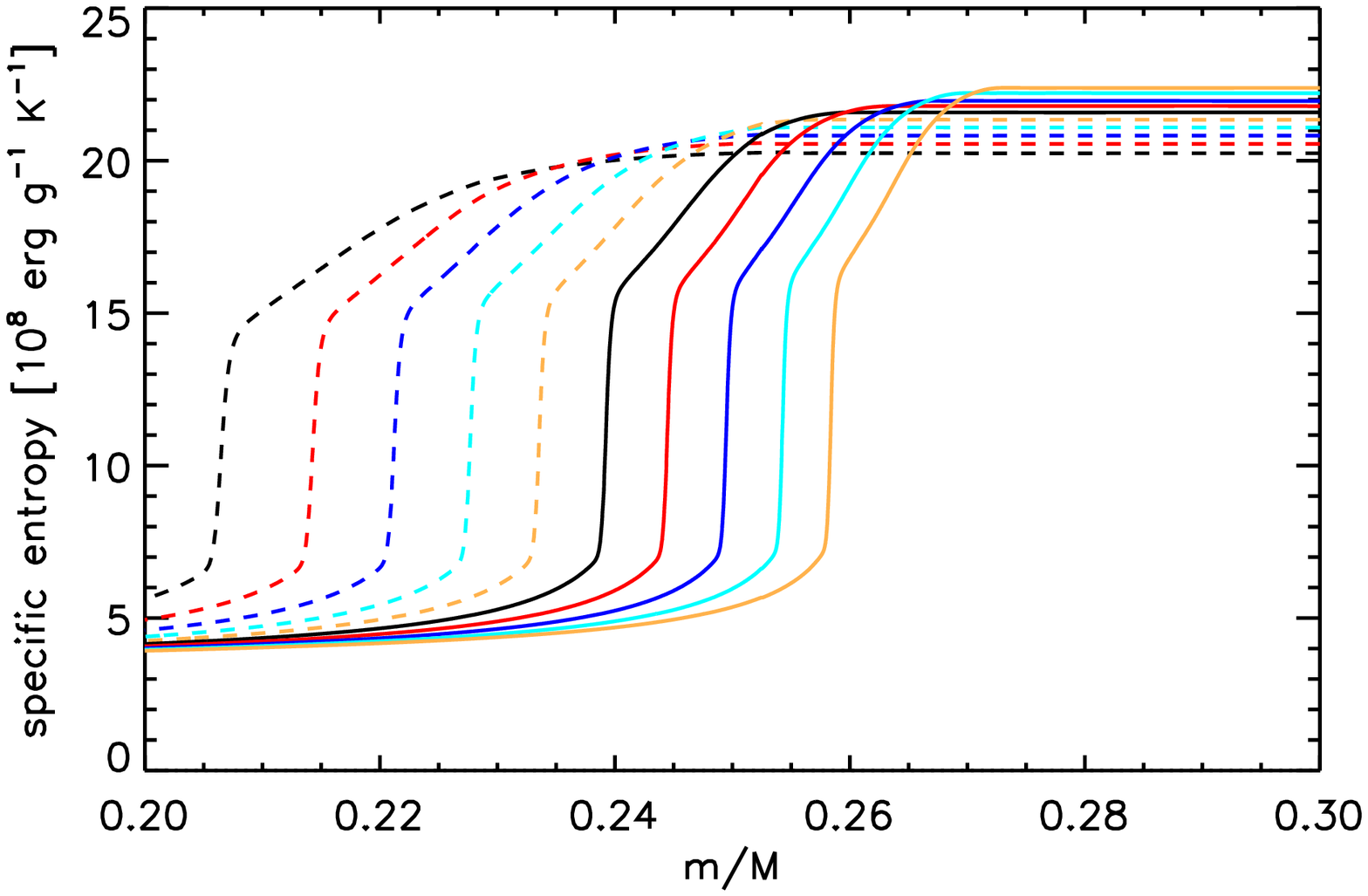}
\end{minipage}
\begin{minipage}{0.9\linewidth}
\includegraphics[width=\linewidth]{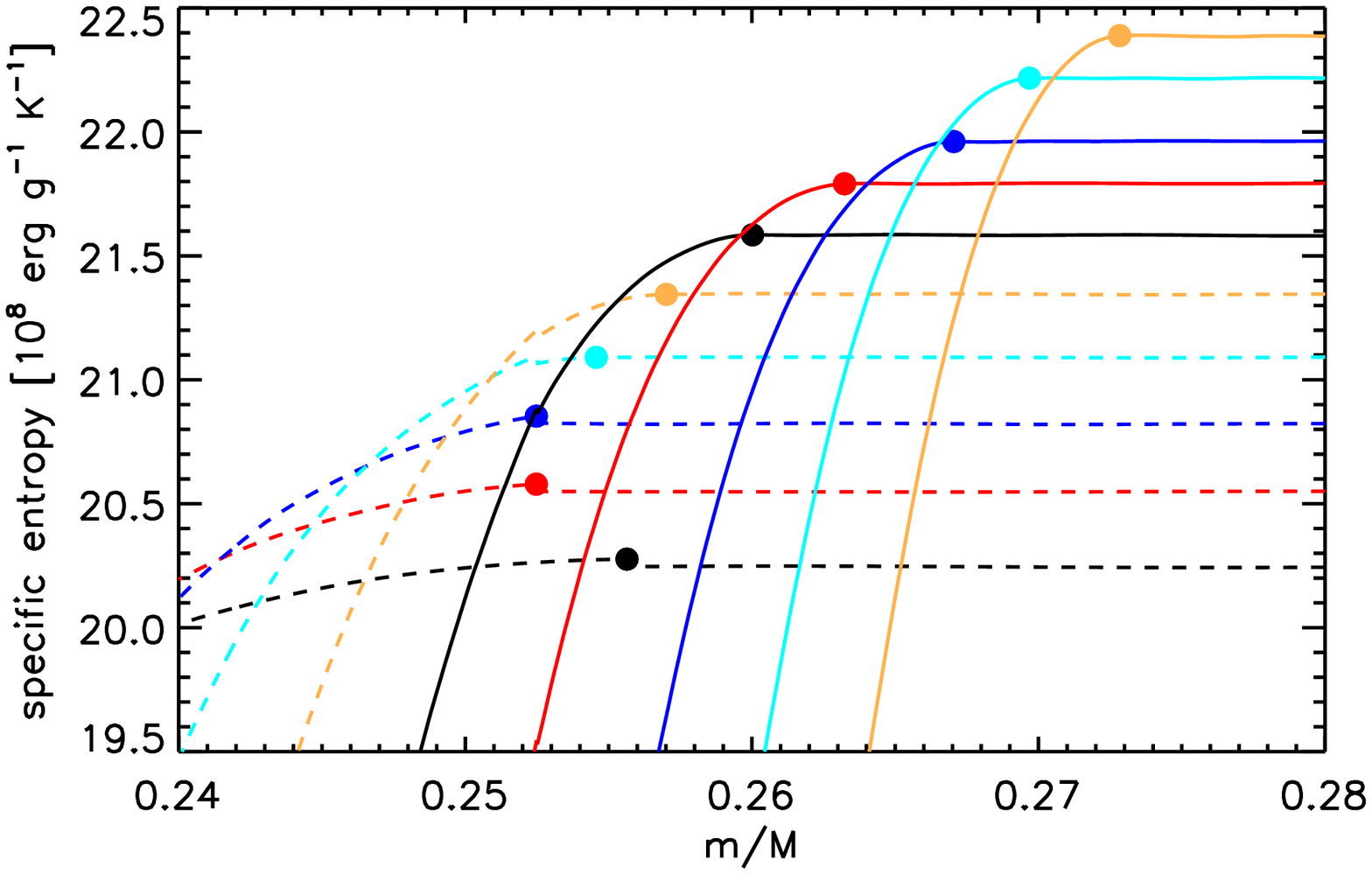}
\end{minipage}
\caption{$\ln(T^{5/2}/P)$ (\textsc{top}), specific entropy (\textsc{middle}) and a zoom of the specific entropy values (\textsc{bottom}) close to the base of the convection zone as a function of mass ordinate for models on a stellar evolution track without a mean molecular weight discontinuity. The dots indicate the location of the base of the convection zone. } 
\label{nodmu}
\end{figure}

In Fig.~\ref{nodmu}, we show $\ln(T^{5/2}/P)$ and specific entropy for the modified models for the parameter range where we would have expected the luminosity bump. We find a monotonic decrease in the minimum of the $\ln(T^{5/2}/P)$ profile and a monotonic increase in the maximum of the $\ln(T^{5/2}/P)$ profile as the star evolves. This is very significant for the formation of a pivot. As is illustrated in the figure, the profile for an older model will always cross the profile of the previous, younger model. Applying Eq.~\ref{sstar}, we see that this implies a continuously decreasing specific entropy in the core and an increasing specific entropy in the convection zone (lower panels of Fig.~\ref{nodmu}) and thus a pivot in all models. 

\begin{figure}
\centering
\begin{minipage}{\linewidth}
\includegraphics[width=\linewidth]{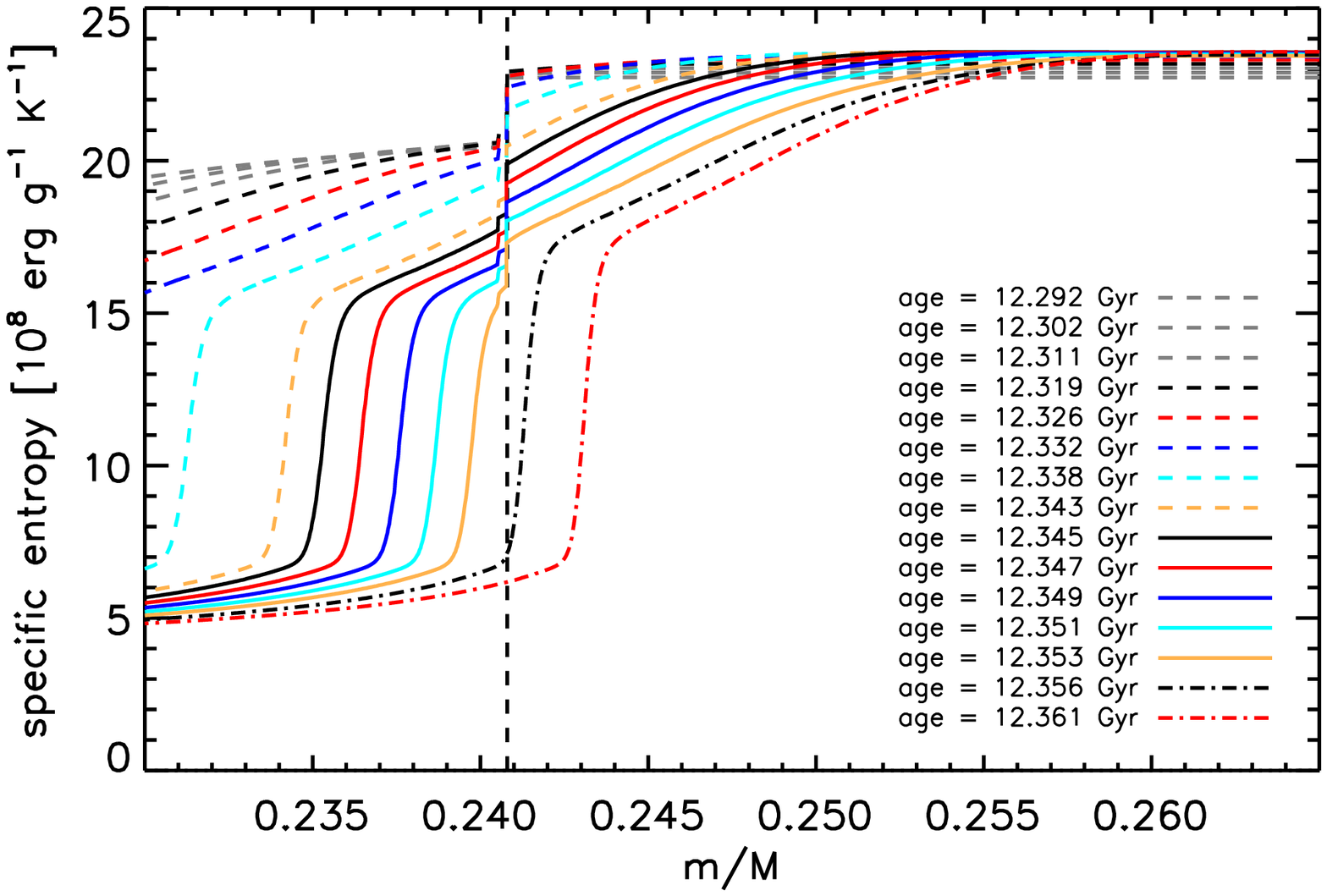}
\end{minipage}
\begin{minipage}{\linewidth}
\includegraphics[width=\linewidth]{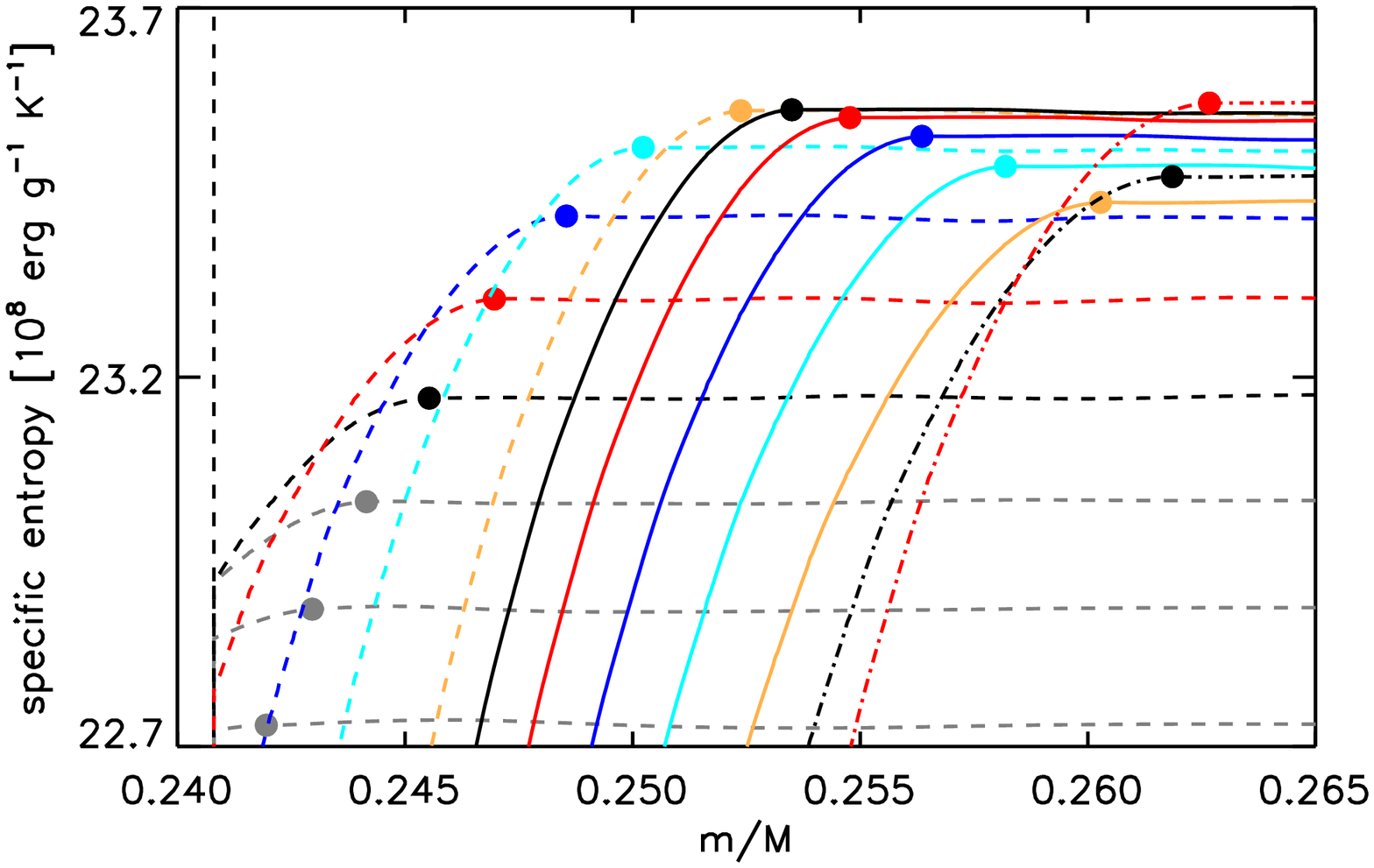}
\end{minipage}
\caption{\textsc{top}: specific entropy of some models around the bump for a standard stellar evolution track with a mean molecular weight discontinuity. The ages are indicated in the legend. Models which occur on the track before the luminosity maximum, between the luminosity maximum and luminosity minimum and after the luminosity minimum are indicated with dashed, solid and dashed-dotted lines, respectively. The vertical dashed line indicates the location in fractional mass of the mean molecular weight discontinuity. \textsc{bottom}: a zoom of the specific entropy values close to the base of the convection zone indicated with the dots.}
\label{dmu}
\end{figure}

We now turn to the specific entropy of models in which a mean molecular weight discontinuity is present (see Fig.~\ref{dmu}). We find that the profiles of the specific entropy are distorted by the step in the mean molecular weight. As the convection zone grows the mean molecular weight discontinuity is at the base of the convection zone. As the convection zone starts receding an entropy difference between the base of the convection zone and the mean molecular weight discontinuity starts to develop. The lower specific entropy values at the mean molecular weight discontinuity (see top panel of Fig.~\ref{dmu}) also lead to reduced specific entropy values at the base of the convection zone (see bottom panel of Fig.~\ref{dmu}) and the star loses its pivot.

To further understand why a star loses its pivot, we develop a toy model in which we aim to disentangle the effect of the mean molecular weight discontinuity from changes due to stellar evolution without a mean molecular weight discontinuity. To achieve this, we compute contributions to the entropy in different regions and sum them to reveal the entropy at the base of the convection zone and how this changes over time. In practice, we rewrite Eq.~\ref{sstar} replacing $N_{\rm A} k_{\rm B}/\mu$ by $x$ and $\ln(T^{5/2}/P)$ by $y$ to become $s=x y$. In this toy model we assume that $y$ follows the evolution of the star without a mean molecular weight discontinuity as shown in the top panel of Fig.~\ref{nodmu}. Furthermore, we assume that at the hydrogen-burning shell, there is a step-like change in $x$ which we denote by $\delta x$, with $x=x_1$ below the step and $x=x_2$ above the step. This is the only step in $x$ that we include in our consideration for now. Owing to the continuously increasing specific entropy from the core to the base of the convection zone (see Eq.~\ref{dsdr}), the value of the entropy at the base of the convection zone $s_{\rm bcz}$ can be approximated to consist of the following contributions:
\begin{equation}
\begin{aligned}
s_{\rm bcz} &= x_1\cdot \delta y_1 \\ &+  \delta x \cdot y|_{\textrm{at H-burning shell}}\\ &+ x_2 \cdot \delta y_2,
\end{aligned}
\label{sbcz}
\end{equation}
where $\delta y_1 =  y|_{\textrm{at H-burning shell}} - y|_{\textrm{m/M = 0}}$ and $\delta y_2 = y|_{\textrm{bcz}} - y|_{\textrm{at H-burning shell}}$.
A change in $s$ at the base of the convection zone per unit time ($\Delta s_{\rm bcz} / \Delta t$) can then be expressed as:
\begin{equation}
\begin{aligned}
\frac{\Delta s_{\rm bcz}}{\Delta t}=\frac{\Delta(xy)}{\Delta t}& = A + B + C ,
\label{deltas}
\end{aligned}
\end{equation}
where 
\begin{eqnarray}
\begin{aligned}
A &= \frac{x_1 \Delta (\delta y_1)}{\Delta t} \\ 
B &= \frac{\delta x \Delta y}{\Delta t} \bigg |_{\textrm{at H-burning shell}}\\
C &= \frac{x_2 \Delta (\delta y_2)}{\Delta t},
\end{aligned}
\end{eqnarray}
and $x_1$, $\delta x$ and $x_2$ are assumed to have values that are constant in time.
The $x$-step leads to a step in the specific entropy located at the hydrogen-burning shell which advances outwards in mass as the star evolves. There is additionally a contribution from the rate of change in $y$ in the regions of constant $x$. These effects are evident as the steep rises in specific entropy at the hydrogen-burning shell and shallower trends elsewhere, as shown in the middle panel of Fig.~\ref{nodmu}. 

\begin{figure}
\centering
\begin{minipage}{\linewidth}
\includegraphics[width=\linewidth]{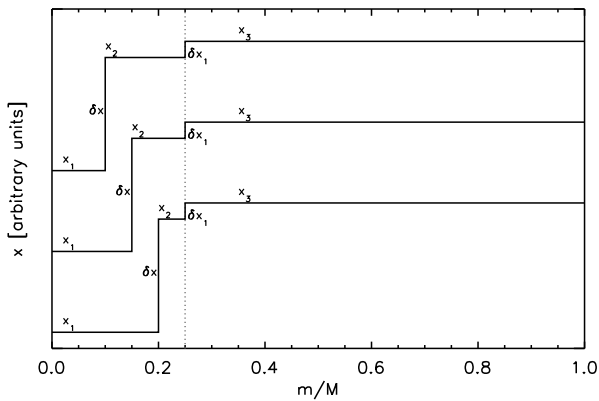}
\end{minipage}
\caption{A schematic illustration of $x$ at three different times evolving from top to bottom (the curves are offset by an arbitrary value for visual purposes). The fractional mass at which the step due to the mean molecular weight discontinuity ($\delta x_1$) occurs is indicated by the vertical grey dotted line. The (larger) step in $x$ at the hydrogen-burning shell ($\delta x$) moves to larger fractional mass as the star evolves.}
\label{step}
\end{figure}

We now extend this toy model to also include a step in the mean molecular weight in the region outside the core (see Fig.~\ref{step}), with a step of height $\delta x_1$ to go from $x=x_2$ below the mean molecular weight discontinuity to $x=x_3$ above the mean molecular weight discontinuity. We note here that unlike the hydrogen-burning shell, the mean molecular weight discontinuity is located at a constant fractional mass after the base of the convection zone has started receding. So as long as the mean molecular weight discontinuity and the hydrogen-burning shell are not co-located, the specific entropy at the base of the convection zone consists of the following contributions:
\begin{equation}
\begin{aligned}
s_{\rm bcz} &= x_1\cdot \delta y_1 \\ &+  \delta x \cdot y|_{\textrm{at H-burning shell}}\\ &+ x_2 \cdot \delta y_2\\ &+  \delta x_1 \cdot y|_{\textrm{at $\mu$-discontinuity}}\\ &+ x_3 \cdot \delta y_3,
\end{aligned}
\label{sbcz_mudisc}
\end{equation}
where $\delta y_2 = y|_{\textrm{at $\mu$-discontinuity}} - y|_{\textrm{at H-burning shell}}$ and $\delta y_3 = y|_{\textrm{bcz}} - y|_{\textrm{at $\mu$-discontinuity}}$.
We can subsequently express the change in specific entropy at the base of the convection zone as:
\begin{equation}
\begin{aligned}
\frac{\Delta s_{\rm bcz}}{\Delta t}=\frac{\Delta(xy)}{\Delta t}& =A + B+ C + D +E, 
\label{deltas_mudisc}
\end{aligned}
\end{equation}
where 
\begin{eqnarray}
\begin{aligned}
D &= \frac{\delta x_1 \Delta y}{\Delta t} \bigg |_{\textrm{at $\mu$-discontinuity}}\\
E &= \frac{x_3 \Delta (\delta y_3)}{\Delta t} 
\end{aligned}
\end{eqnarray}
and $x_1$, $\delta x$, $x_2$, $\delta x_1$, $x_3$ are assumed to have values that are constant in time.

Initially, as $\delta x_1$ is still located at or close to the base of the convection zone the value of $y$ at the location of $\delta x_1$ is large and changes slowly, i.e. $\Delta y / \Delta t$ at $\delta x_1$ is small. Then, as the hydrogen-burning shell encroaches on the mean molecular weight discontinuity and the base of the convection zone recedes, the $y$-value at $\delta x_1$ starts to decrease, causing a negative contribution of $D$ to $\Delta s_{\rm bcz}/\Delta t$.

\begin{figure*}
\centering
\begin{minipage}{0.48\linewidth}
\includegraphics[width=\linewidth]{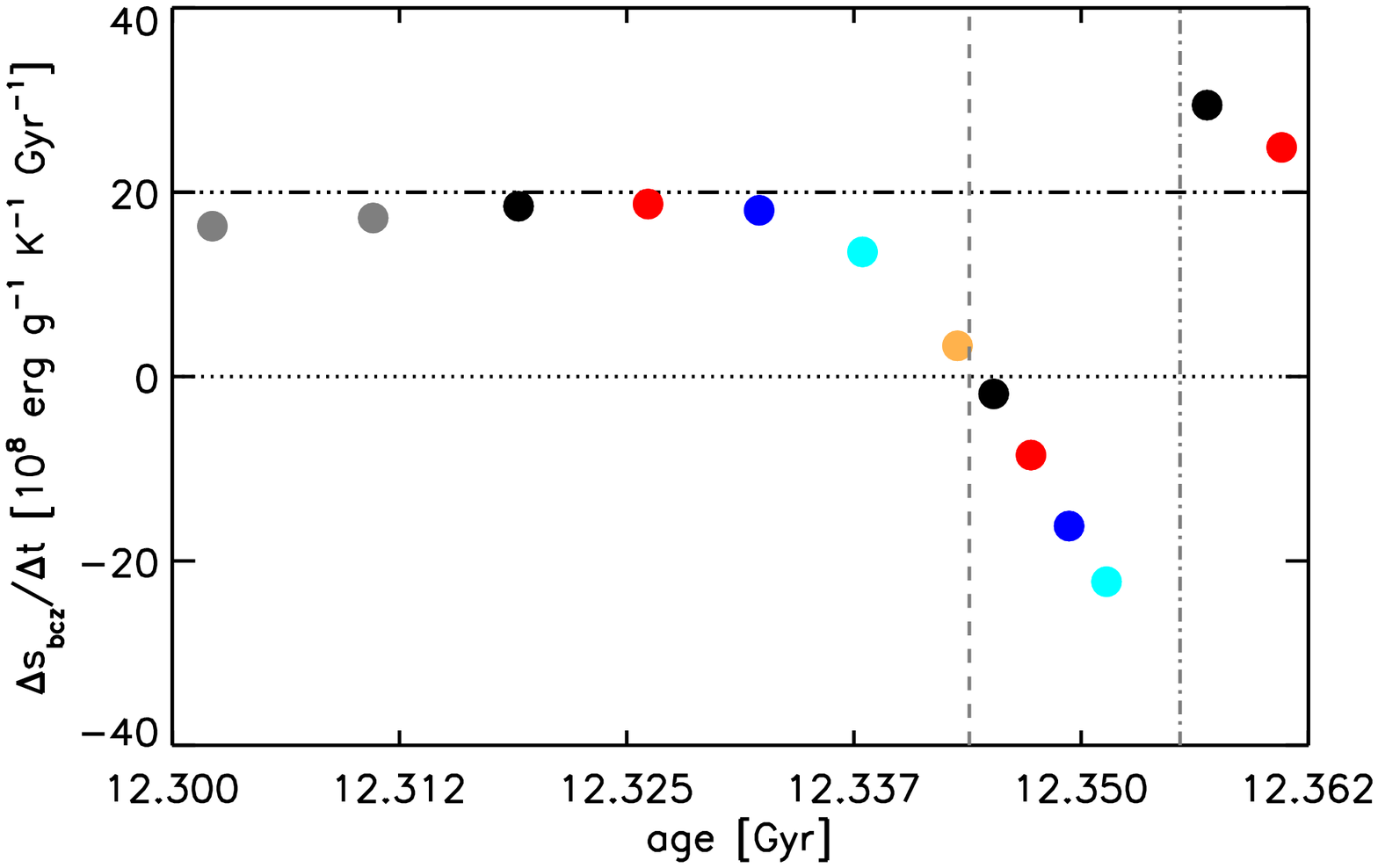}
\end{minipage}
\begin{minipage}{0.48\linewidth}
\includegraphics[width=\linewidth]{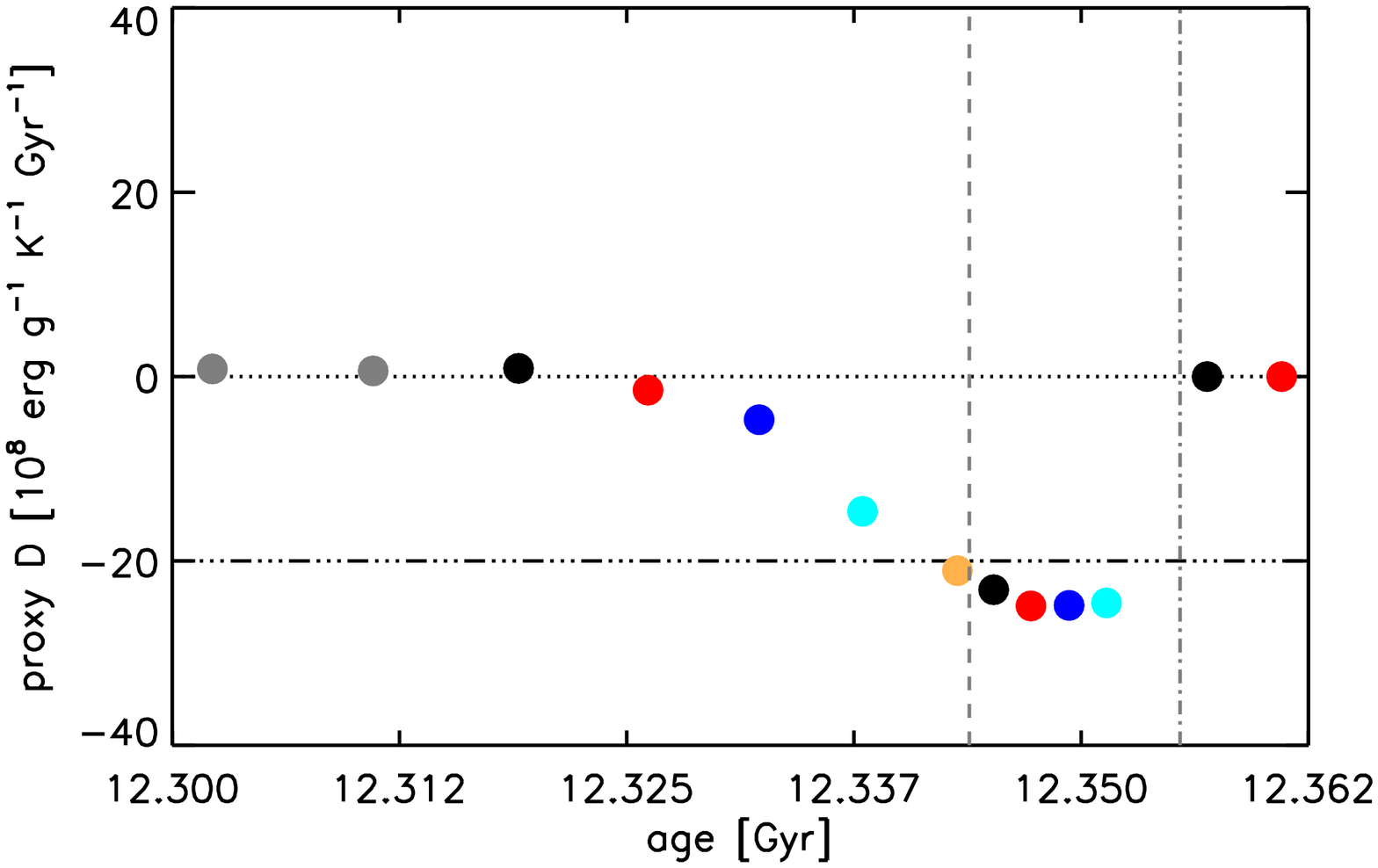}
\end{minipage}
\caption{\textsc{left}: Change in specific entropy at the base of the convection zone per unit time ($\Delta s_{\rm bcz}/\Delta t$, as per Eq.~\ref{deltas_mudisc}) as a function of age. The slowly increasing trend before 12.325\,Gyr changes to a decreasing trend due to the specific entropy beginning to sense the approach of the mean molecular weight discontinuity. Note that this behaviour is in line with the values of $\epsilon_{\rm g}$ as shown in Fig.~\ref{bcz}. The horizontal dotted line indicates zero. The vertical dashed and dashed-dotted lines indicate the stellar ages of the luminosity maximum and luminosity minimum of the bump, respectively. The horizontal dashed-triple-dotted line indicates a value of $20\cdot10^8$\,erg\,g$^{-1}$\,K$^{-1}$\,Gyr$^{-1}$.\newline \textsc{right}: The change in the step in specific entropy at the mean molecular weight discontinuity as a function of age. This quantity is a proxy for $D$. In this case the horizontal triple-dotted-dashed line indicates a value of $-20\cdot10^8$\,erg\,g$^{-1}$\,K$^{-1}$\,Gyr$^{-1}$. The other lines have the same meaning as in the left panel.}
\label{dds}
\end{figure*}

To verify if the decrease in $D$ could indeed counter-act the other contributions, we look at the change of the specific entropy at the base of the convection per unit time $\Delta s_{\rm bcz} / \Delta t$ in the left panel of Fig.~\ref{dds}, and at the change in the step in specific entropy at the mean molecular weight discontinuity which serves as a proxy for $D$ (right panel of Fig.~\ref{dds}). We find that for the models prior to the luminosity maximum of the bump  $\Delta s_{\rm bcz} / \Delta t$ first increases slowly to a value of about $20\cdot10^8$\,erg\,g$^{-1}$\,K$^{-1}$\,Gyr$^{-1}$ while the contribution from $D$ is close to zero. As the base of the convection zone recedes further and the hydrogen-burning shell approaches the mean molecular weight discontinuity, $D$ is negative and decreases to values below $-20\cdot10^8$\,erg\,g$^{-1}$\,K$^{-1}$\,Gyr$^{-1}$, indeed counter-acting the positive contributions from the other parts of the star to $\Delta s_{\rm bcz} / \Delta t$ as indicated in the right-hand side of Eq.~\ref{deltas_mudisc}. Thus, the resulting $\Delta s/ \Delta t$ is negative throughout the star, i.e. $\epsilon_{\rm g}$ is positive and no pivot exists.

With this toy model, we show that the approach of the hydrogen-burning shell towards the mean molecular weight discontinuity is sensed by the specific entropy. Away from the discontinuity, at each stage in the evolution, the specific entropy at fractional masses beyond the hydrogen-burning shell continues to increase causing the specific entropy curves to cross and the formation of a pivot below the convection zone. As stated earlier, no pivot can occur within the convection zone. Beyond the hydrogen-burning shell, the specific entropy slowly increases up to the base of the convection zone. Both the hydrogen-burning shell and the base of the convection zone move to higher fractional mass as the star evolves, while the discontinuity remains at fixed mass fraction. Eventually, as the mean molecular weight discontinuity is approached by the hydrogen-burning shell the situation changes. The discontinuity co-incides with decreasing values of the specific entropy. We have shown that this reduces the temporal gradient in the specific entropy and leads to the disappearance of the pivot. Without a pivot the star is fully contracting and decreases in luminosity. This explains the luminosity maximum of the bump. The process is terminated when the hydrogen-burning shell reaches the mean molecular weight discontinuity and removes it.

\section{Discussion and Conclusions}
In this study we explored the cause of the luminosity maximum of the RGB bump. We did so by exploring changes in the specific entropy as the star evolves. Temporal changes of the specific entropy have been addressed through studying the gravothermal energy generation rate $\epsilon_{\rm g}$. We defined the fractional mass at which $\epsilon_{\rm g}$ changes from being positive in the core to negative in the outer layers as the `pivot'. The stationary point, i.e. the mass fraction at which $\partial r/\partial t =0$, is located at larger fractional mass than the pivot. The stationary point indicates the division between inward moving mass shells (in the deep interior) and outward moving mass shells above the stationary point. We introduced the pivot and the stationary point as well-defined attributes of the mirror phenomenon.

As the star evolves, we find that the stationary point crosses the inwardly moving base of the convection zone (red crossing blue in the top panels of Fig.~\ref{location}). This means that the mass shells just above the base of the convection zone reverse their direction of movement. Since $\epsilon_{\rm g}$ remains negative, these layers continue to decrease in density. As the stars keeps evolving these layers with decreasing density start to move inwards. We speculate that this may trigger the recession of the base of the convection zone. We will address this in a future study.

We also find that at some later point in time the pivot reaches the base of the convection zone, and both the pivot and the stationary point disappear; the star does not have a mirror anymore and is fully contracting. The disappearance of the pivot coincides with the luminosity maximum of the bump. The star regains a pivot when the hydrogen-burning shell reaches the mean molecular weight discontinuity. In line with \citet{JCD2015}, who has already shown that the encounter of the hydrogen-burning shell with the mean molecular weight discontinuity coincides with the luminosity minimum of the bump. 

We have conducted a detailed investigation into the cause of the disappearance of the pivot at the luminosity maximum of the bump. As the hydrogen-burning shell approaches the mean molecular weight discontinuity, we find that the discontinuity is at layers with lower and lower entropy. Using a toy model, we find that with lower values of specific entropy at the mean molecular weight discontinuity, the step in the specific entropy caused by the mean molecular weight discontinuity decreases. This decrease counter-acts the temporal increase in specific entropy at the base of the convection zone predominantly due to the temperature-pressure ratio ($T^{5/2}/P$), and thereby removes the pivot.
To verify the role of the mean molecular weight discontinuity, we also computed a stellar evolution track in which the mean molecular weight discontinuity was removed. Indeed, a pivot is present in all models along this track and no bump feature appears.

We note that others have investigated the impact of the size and shape of the mean molecular weight discontinuity on the bump features. In particular \citet{cassisi2002} study the effect of smoothing the mean molecular weight discontinuity. They find that in stellar models whose mean molecular weight discontinuity has been smoothed, the drop in luminosity is correlated with the smoothing length. They also investigated the change in opacity profile caused by variation in the chemical stratification and demonstrated that it has a negligible effect on the RGB bump. Hence, they conclude that the change in mean molecular weight at the discontinuity determines the shape of the RGB bump. In line with these results, we also find that the step size in the mean molecular weight discontinuity (as described by $\delta x_1$ in Eqs~\ref{sbcz_mudisc} \& \ref{deltas_mudisc} and Fig.~\ref{step}) is a critical parameter of the RGB bump.

To distinguish in what phase of the bump a particular star is both \citet{townsend2013} and \citet{gai2015} have investigated the behaviour of oscillations in models passing through the bump. \citet{townsend2013} predict a temporary increase in the otherwise-decreasing frequencies of the avoided crossings. Similarly, \citet{gai2015} find that the bump is visible in the dipole period spacing.

Finally, current models show a mismatch with observations \citep[e.g.][]{khan2018}, and we speculate that the physical principles that we have employed in this work may be of use to improve the theoretical predictions of the luminosities of the RGB bump. 

\section*{Acknowledgements}
We thank J. Christensen-Dalsgaard for useful comments on earlier versions of this manuscript that improved the manuscript significantly.
The research leading to the presented results has received funding from the European Research Council under the European Community's Seventh Framework Programme (FP7/2007-2013) / ERC grant agreement no 338251 (StellarAges). YE acknowledges the support of the UK Science and Technology Facilities Council (STFC). SB acknowledges NASA grant NNX16AI09G.
Funding for the Stellar Astrophysics Centre (SAC) is provided by The Danish National Research Foundation (Grant agreement no.: DNRF106).




\bibliographystyle{mnras}
\bibliography{entropyv1} 

\bsp	
\label{lastpage}
\end{document}